\documentstyle[12pt]{article}
\input psfig

\def\Mpc{\,{\rm Mpc}}

\def\cmm2{{\,\rm cm^{-2}}}
\def\cm2{{\,{\rm cm}^2}}
\def\cmm3{{\,{\rm cm}^{-3}}}
\def\gcmm3{{\,{\rm g\,cm^{-3}}}}

\def\mpl{{m_{\rm Pl}}}

\def\fun#1#2{\lower3.6pt\vbox{\baselineskip0pt\lineskip.9pt
  \ialign{$\mathsurround=0pt#1\hfil##\hfil$\crcr#2\crcr\sim\crcr}}}

\begin{document}
\baselineskip=14pt
\pagestyle{empty}
\begin{center}
\bigskip


\vspace{.2in}
{\Large \bf KINEMATIC CONSTRAINTS TO THE \\
\medskip
KEY INFLATIONARY OBSERVABLES}
\bigskip

\vspace{.2in}
Mark B. Hoffman$^1$ and Michael S. Turner$^{1,2,3}$\\

\vspace{.2in}

{\it $^1$Department of Physics\\
The University of Chicago, Chicago, IL~~60637-1433}\\

\vspace{0.1in}
{\it $^2$Department of Astronomy \& Astrophysics\\
Enrico Fermi Institute, The University of Chicago, Chicago, IL~~60637-1433}\\

\vspace{0.1in}
{\it $^3$NASA/Fermilab Astrophysics Center\\
Fermi National Accelerator Laboratory, Batavia, IL~~60510-0500}\\

\end{center}

\vspace{.3in}
\centerline{\bf ABSTRACT}

The observables $T/S$ and $n-1$ are key to testing and understanding inflation.
($T$, $S$, and $n-1$ respectively quantify the gravity-wave
and density-perturbation contributions to CMB anisotropy and the deviation
of the density perturbations from the scale-invariant form.)
Absent a standard model, there is no definite prediction for,
or relation between, $T/S$ and $n-1$.
By reformulating the equations governing inflation
we show that models generally predict $T/S\approx -5(n-1)$ or $0$, and
in particular, if $n>0.85$, $T/S$ is expected to be $>10^{-3}$.

\bigskip

\newpage
\pagestyle{plain}
\setcounter{page}{1}
\newpage

{\em Introduction.}  Cosmic microwave background (CMB) anisotropy measurements
have begun to test inflation, the leading paradigm to extend the
standard big-bang cosmology.
Within a decade they should test inflation decisively and even
probe the underlying physics \cite{MAP,Planck,Hu}.
Recent results from the BOOMERanG and MAXIMA CMB experiments \cite{Boom,maxima}
(as well as results from earlier experiments \cite{knox-page})
confirm the flat Universe predicted by inflation
and are beginning to address its second basic prediction:
almost scale-invariant adiabatic, Gaussian density perturbations
produced by quantum fluctuations during inflation \cite{scalar}.
The third prediction, a nearly scale-invariant spectrum of gravity
waves, will be more
difficult to confirm, but is a critical probe of inflation \cite{gw}.

The key inflationary observables are:  the level of anisotropy arising from
density (scalar) perturbations (quantified by the contribution to the
CMB quadrupole anisotropy, $S$), the level of anisotropy arising from
gravity-wave (tensor) perturbations
($T$), and the power-law index $n$ that characterizes the density perturbations
(scale invariance refers to equal amplitude fluctuations
in the gravitational potential on all length scales and corresponds to $n=1$).
If $T$, $S$ and $n-1$ can be measured, then
the scalar-field potential that drove
inflation can be reconstructed \cite{recon}.
The most promising means of measuring $T$ is its unique signature
in the polarization of CMB anisotropy \cite{CMB_pol} (however, direct
detection by a future space-based experiment should not
be dismissed).

While there is no standard model of inflation,
all known models can be cast in terms of the classical evolution
of a new scalar field $\phi$ (dubbed the inflaton)
\cite{lyth-riotto}.  Predictions for $S$, $T$ and $n-1$ can be
expressed in terms of the scalar-field potential $V(\phi $)
and its first two derivatives.  While there is a model-independent relation
between $T/S$ and the power-law index $n_T$ that characterizes the gravity-wave
spectrum, $T/S = -5n_T$ \cite{consistency,note1},
no such relation for $n$ and $T/S$ exists \cite{davisetal}.

This is unfortunate because $n_T$ is very
difficult to measure, and $n$ will be measured to
a precision of better than 1\% by the MAP and PLANCK
experiments (BOOMERanG and MAXIMA have already
determined that $n =1\pm 0.06$).  Even an approximate or generic
relation between $(n-1)$ and $T/S$ would be valuable, both as a
test of inflation and as a guide for the expected level of
gravity waves when $n$ is measured.

The formation of large-scale structure and CMB measurements
already indicate that a significant part of CMB anisotropy
arises from scalar perturbations ($T/S$ cannot be $\gg 1$).  On the
other hand, nothing precludes $T/S \ll 1$, and if $T/S$ is much
less than $10^{-3}$, the prospects for measuring $T$
are poor \cite{CMB_pol}
(one inflation theorist has opined that $T/S \ll 1$
for all reasonable models \cite{lyth}).

The goal of this work is to provide objective
theoretical guidance.  By casting the equations
governing inflation in a form that is essentially independent
of the inflaton potential (``flow equations'' for $T/S$ and
$n-1$), we show that
the $T/S$ -- $(n-1)$ plane is not uniformly populated by models of inflation:
The lines $T/S \approx 0$ and $T/S \approx -5(n-1)$ act as attractors
for models that are consistent with the equations governing inflation.
For $n<1$, there is an excluded region between these two
attractors; for $n>1$, other values for $T/S$ and $n-1$ are
possible, but at the expense of a spectrum of density perturbations
that is poorly represented by a power law.  (the CMB will
be able to test how well a power law describes the density perturbations.)

{\em Flow equations.} The kinematic equations that govern inflation are
well known \cite{KT}
\begin{eqnarray}
\ddot{\phi} + 3H\dot{\phi} + V^\prime (\phi) & = & 0 \\
H^2 \equiv \left( \frac{\dot{a}}{a} \right)^2 & = &
  \frac{8\pi}{3\mpl^2} \left[ V(\phi) + \frac{1}{2}\dot{\phi}^2\right]
\end{eqnarray}
where $a(t)$ is the cosmic scale factor, prime denotes $d/d\phi$,
and overdot denotes $d/dt$.  During inflation $\phi$ rolls slowly
and the $\ddot{\phi}$ term in
its equation of motion and its kinetic term
in the Friedmann equation can be neglected \cite{KT,ST}, so that
\begin{eqnarray}
\label{sr1}
\dot{\phi} & \simeq & \frac{-V^\prime}{3H} \\
\label{sr2}
N(\phi ) \equiv \int^{\phi_{\rm end}}_\phi Hdt
& \simeq & -\frac{8\pi}{\mpl^2} \int_{\phi}^{\phi_{\rm end}}
\frac{d\phi}{x(\phi)}
\end{eqnarray}
where $x(\phi) \equiv V^\prime (\phi)/V(\phi)$ measures the steepness of
the potential and $N(\phi )$, the number of e-folds before the
end of inflation, is the natural time variable.
Inflation ends when the slow-roll conditions,
\begin{eqnarray}
\label{srlim1}
\mpl V^\prime /V = \mpl x & < & \sqrt{48\pi} \\
\label{srlim2}
\mpl ^2 V^{\prime\prime}/V = \mpl^2 (x^\prime +x^2) & < & 24\pi\,,
\end{eqnarray}
are violated (at $\phi = \phi_{\rm end}$) \cite{KT,ST}.

The inflationary observables are related to the same quantities that
govern the kinematics of inflation \cite{note1}
\begin{eqnarray}
\label{n1}
(n-1) & = & \frac{\mpl ^2}{8\pi} \left[ 2x' - x^2 \right] \\
\label{r}
T/S & = & \frac{5\mpl ^2}{8\pi} x^2 \\
\label{T}
T & = & 0.6 V/\mpl^4
\end{eqnarray}
These expressions are given to lowest order in $x^2$ and $x'$
(see Ref.~\cite{LT} for higher-order corrections).  Note, $n-1$
is only equal to $n_T =-5(T/S)$ if $x^\prime = 0$.

By combining the slow-roll equations with those governing
$(n-1)$ and $T/S$, we can write equations that govern
the inflationary observables (almost) without reference to a model,
\begin{eqnarray}
\label{flowr}
\frac{d(T/S)}{dN} & = & (n-1)\frac{T}{S} +
		\frac{1}{5}\left(\frac{T}{S} \right) ^2 \\
\label{flown}
\frac{d(n-1)}{dN} & = &
   - \frac{1}{5}(n-1)\frac{T}{S} - \frac{1}{25}\left(\frac{T}{S}\right) ^2
   \pm \frac{\mpl ^3}{16\pi^2}\sqrt{\frac{2\pi}{5}\frac{T}{S}}\,x^{\prime\prime}
\end{eqnarray}
where the sign of the last term matches that of $V^\prime$.

We call these ``flow equations''
as they describe the trajectory in the $T/S$ --
$(n-1)$ plane during inflation.  Because of the
$x^{\prime\prime}$ term they are not completely independent
of the potential.  To ``close'' the flow equations
we will assume that the potential is smooth enough so that
we can treat $x^{\prime\prime}$ as being approximately constant.
For sufficiently smooth and featureless potentials $x^{\prime\prime}$
will also be small.

Finally, one might wonder what happened to the most stringent
constraint on inflation:  achieving density perturbations of
amplitude $10^{-5}$ or so ($S\sim 10^{-10}$).  The flow equations
involve the quantities $T/S$, $(n-1)$ and $dN/d\phi$ which
are unaffected by a rescaling of the potential, $V\rightarrow aV$.
This rescaling changes $S$:  $S\rightarrow a S$.  Thus, any potential
can be rescaled to give proper size density perturbations without
affecting the flow equations.

{\em Trajectories and attractors.} The scales relevant for
structure formation ($1\Mpc$ to $10^4\Mpc$) crossed outside the
horizon roughly 50 $e$-folds before the end of inflation (i.e.,
when $N =50$) \cite{KT}, and so it is $T/S$ and $(n-1)$ at this time
that can be measured by CMB experiments.
We find them by evolving $T/S$ and $(n-1)$ until inflation
ends and counting back 50 e-folds.
To determine when inflation ends, we recast the slow-roll conditions
(\ref{srlim1},\ref{srlim2}):
\begin{eqnarray}
\label{rlim}
T/S & < & 30 \\
\label{nlim}
\left| (n-1) +\frac{3}{5}\frac{T}{S} \right| & < & 6
\end{eqnarray}

\begin{figure}
\centerline{\psfig{figure=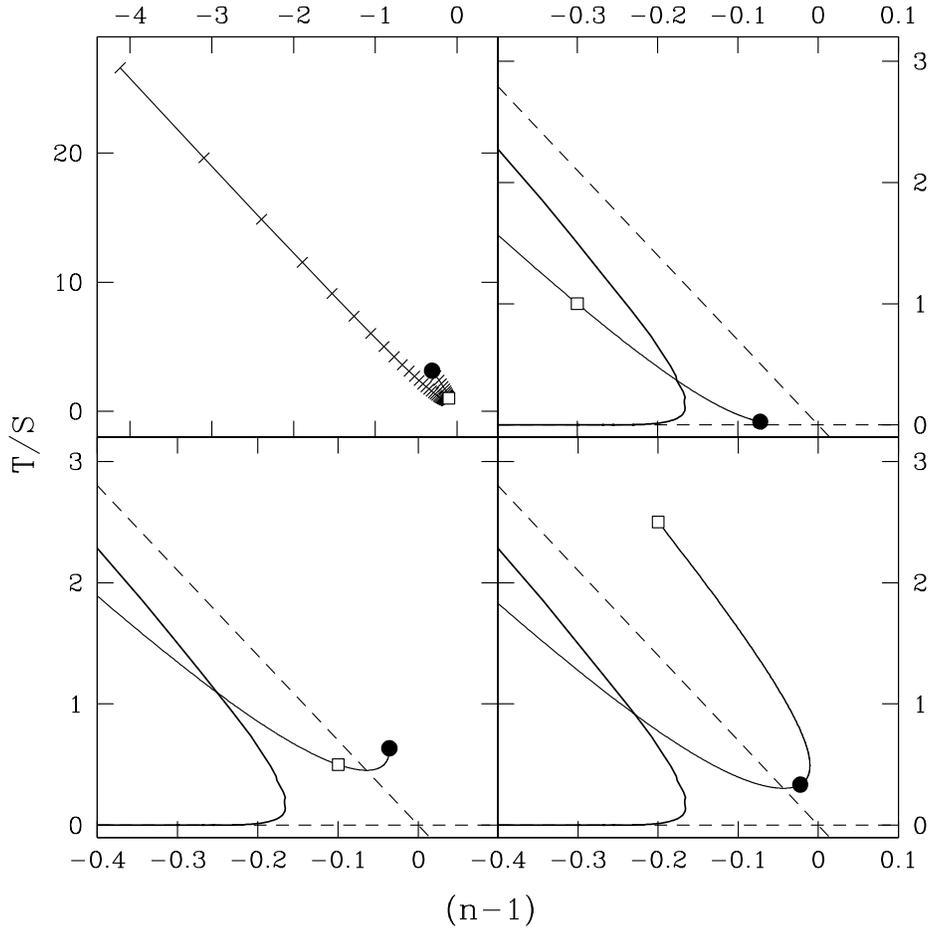,width=5in}}
\caption{Trajectories in the $T/S$ -- $(n-1)$ plane.
Squares indicate the initial choices for $T/S$ and $(n-1)$;
circles indicate the values 50 e-folds before the end of inflation.
A trajectory ends when $T/S$ and/or $|n-1|$ become large;
most of inflation occurs when $T/S$ and $|n-1|$ are small.
The upper left panel shows a complete trajectory, with ticks indicating
e-folds before the end of inflation (from the circle, $50, 49, \cdots , 1$).
The other three panels show trajectories in more detail.
Note how $T/S$ and
$(n-1)$ outside the attractor region are ``pulled in'' (the
attractors are shown as broken lines and the boundary of the
excluded region is a solid curve).
}
\label{fig:tracks}
\end{figure}

To be specific, we pick ``initial'' values in the range,
$0 <  T/S  < 10$ and $-0.5 < (n - 1) < 0.5$,
and then integrate with fixed $x^{\prime\prime}$
until one of the slow-roll conditions
is violated, signaling the end of inflation.
We then count back 50 $e$-folds to find
$(T/S)_{50}$ and $(n-1)_{50}$.  Some trajectories are shown
in Fig.~\ref{fig:tracks}.

\begin{figure}
\centerline{\psfig{figure=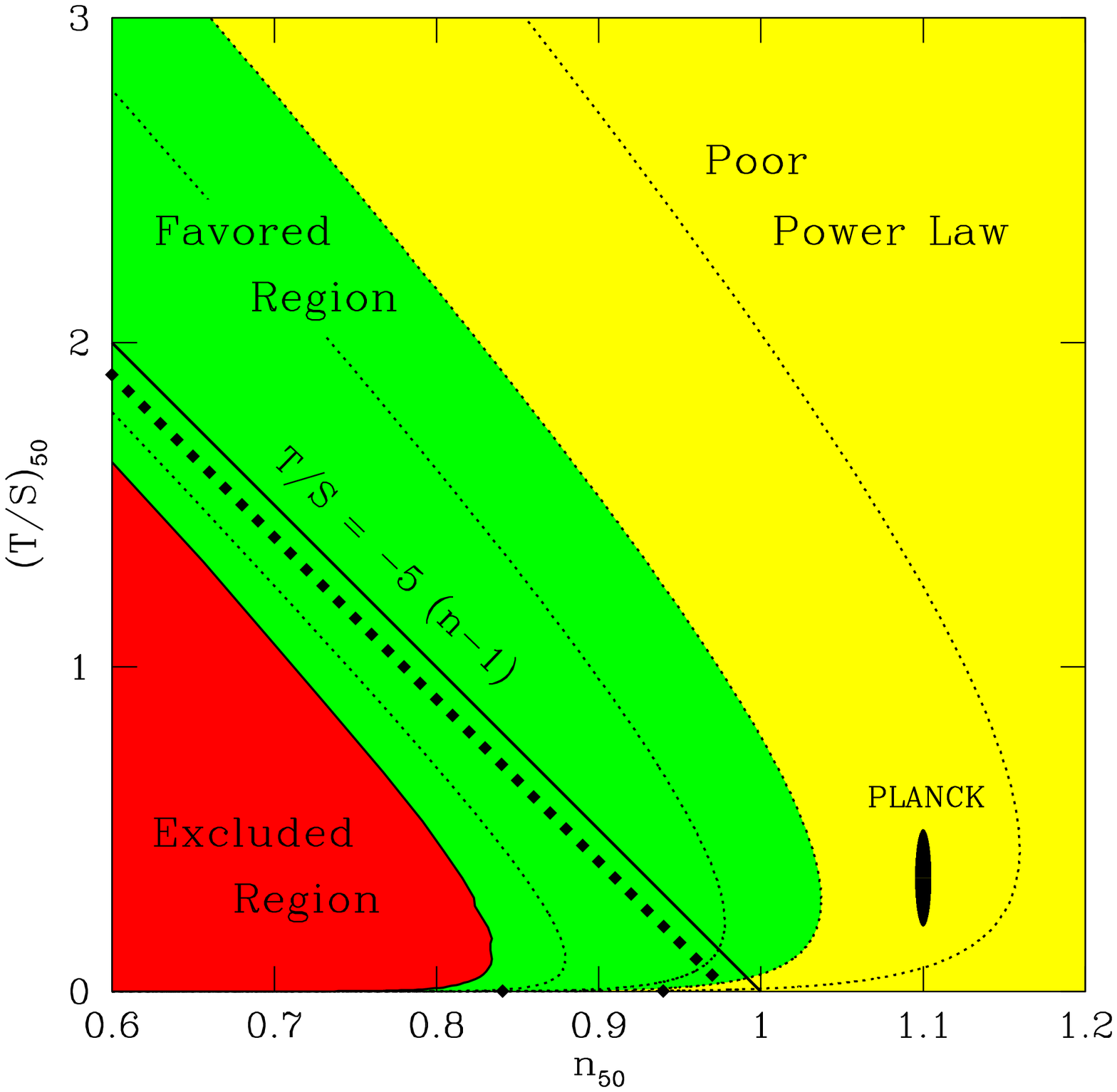,width=5in}}
\caption{Summary of our model search using the flow
equations.  The lines $T/S =-5(n-1)$ and $T/S=0$ act
as attractors; the dotted curves correspond to $x^{\prime\prime}
=0,1,2,5$ (from left to right).  We found no model in the excluded
region, and we call the region between it and the curve $x^{\prime\prime}
=3$ the favored region.  Models outside the favored region
(upper right part) have
large $dn/d\ln k$ and density perturbations that are not well
represented by a power law.  Diamonds indicate various
known inflationary models:  chaotic, $V(\phi )=\lambda\phi^n$ for $n=2,3,\cdots$
(diamonds on the diagonal); new inflation ($n=0.94$) and natural inflation
(with $n=0.84$).  The ellipse is the $2\sigma$ error ellipse ``forecasted''
for the PLANCK satellite \protect\cite{Kinney}.
}
\label{fig:1fld}
\end{figure}

\begin{figure}
\centerline{\psfig{figure=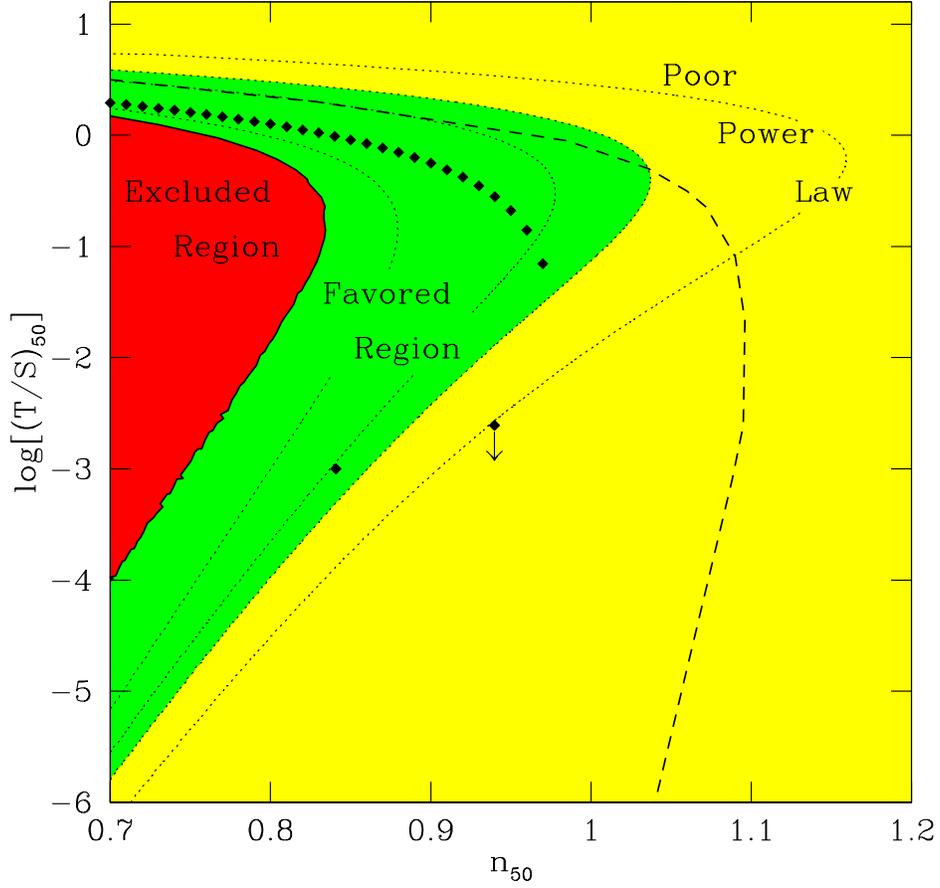,width=5in}}
\caption{Same as Fig.~\protect\ref{fig:1fld}, except with
a logarithmic scale for $T/S$ to show more detail.
As $n\rightarrow 1$, $T/S$ increases in the favored region.
Below and to the left of the broken line ($dn/d\ln k
=10^{-2}$), a poor power law does not occur for large $x^{\prime\prime}$
because $dn/d\ln k \propto \protect\sqrt{T/S}\,x^{\prime\prime}$
and $T/S$ is small.
}
\label{fig:1fldb}
\end{figure}

Figs.~\ref{fig:1fld} and \ref{fig:1fldb} summarize the $(T/S)_{50}$ --
$(n-1)_{50}$ phase space generated from the range of initial conditions
considered.  The $(T/S)_{50}$ -- $(n-1)_{50}$ plane
is not uniformly populated.  For $x^{\prime\prime}< {\cal O}(1)$,
solutions cluster around two attractors, $(T/S)_{50} \approx 0$ and
$(T/S)_{50} \approx -5(n-1)_{50}$, and for $(n-1)_{50} < 0$, there is
an excluded region between the two attractor lines,
which cannot be reached for any value of $x^{\prime\prime}$.
We call the region between the excluded area and
$x^{\prime\prime} = 3$ as the favored region for the inflationary
observables $n-1$ and $T/S$.

Taking $x^{\prime\prime} =0$ it is simple to show how the attractors
arise.  In this limit, the flow equations are:
$s\equiv (n-1)-{1\over 5}{T\over S}
=\,$const and $r \propto \exp (sN)$, where $r=T/S$.  Unless $r$ and
/or $s$ are small, corresponding to the attractor solutions,
$r$ grows very rapidly and inflation does not last 50 e-folds.

For $n>1$, values of $(T/S)$ and $(n-1)$ outside the favored region are
possible at the expense of large $x^{\prime\prime}$.
Models with a large $x^{\prime\prime}$ have density-perturbation spectra
which are not well represented by a power law:  the running of the
spectral index \cite{run_kt}, $d n /d\ln k = -dn/dN$ includes the term
\begin{equation}
{dn \over d\ln k} = \cdots \pm {\mpl^3 \over 16\pi^2}
\sqrt{{2\pi \over 5}{T\over S}}\,x^{\prime\prime}
\end{equation}
which becomes large for large $x^{\prime\prime}$.  This explains
the results of a recent paper \cite{hutererturner} in which models
with $n$ as large as 2 were constructed.  In particular, for the model
with $n=2$, $x^{\prime\prime} \simeq 2000$, $T/S\simeq 3\times 10^{-3}$
and $dn/d\ln k \simeq 0.3$.

Fig.~\ref{fig:1fldb} shows the $T/S \approx 0$ attractor region in
more detail.  As $n$ increases toward unity, values of $T/S$ in
the favored region grow, making the prospects for measuring
$T$ better; in particular, for $n> 0.85$, $T/S > 10^{-3}$.  This
figure also confirms that almost, but
not exactly, scale-invariant density perturbations are
a generic prediction of inflation \cite{scalar,ST}:  the favored region
just touches $n=1$ (for $x^{\prime\prime} \simeq 1.5$).  Finally,
in the disfavored region where $T/S \ll 1$, large $x^{\prime\prime}$
does not imply a poor power law because $dn/d\ln k$ is proportional
to $T/S$.  Indeed one of the models in this region is new inflation.

So far, we have only considered one-field inflation.
There are potentials that are so flat and smooth that the slow-roll conditions
never break down; the most well known of these is power-law inflation,
$V(\phi ) \propto \exp (-\beta \phi / \mpl)$.  In a ``never ending'' model,
another field causes the slow-roll conditions to break down (e.g., by
classical evolution in hybrid inflation or
a phase transition in extended inflation).  The flow equations can also
be applied to such models.

\begin{figure}
\centerline{\psfig{figure=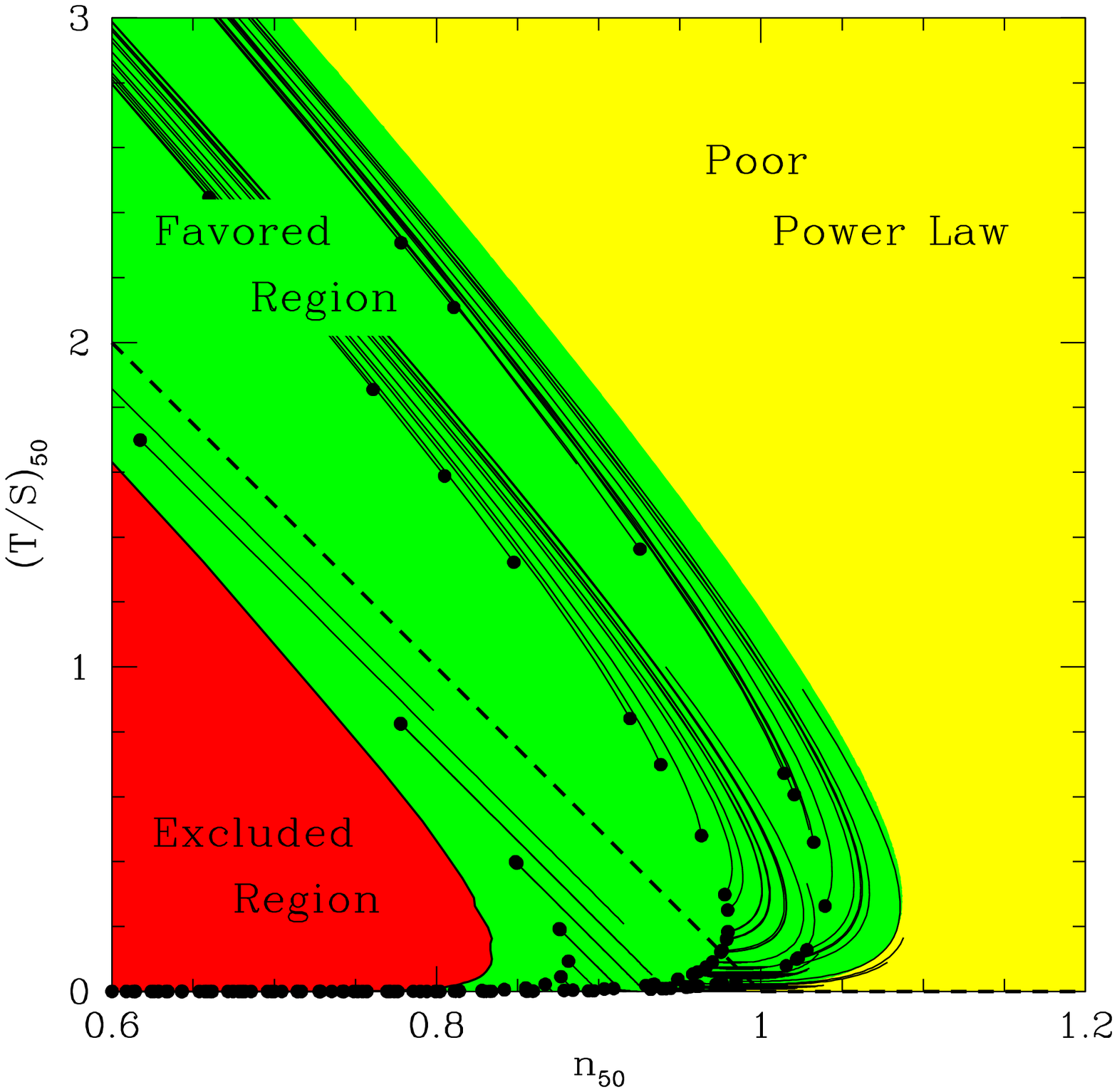,width=5in}}
\caption{Summary of two-field models.  The filled circles represent the
values of $(T/S)_{50}$ and $n_{50}$ for the corresponding one-field
models, and the attached curves are the values obtained for inflation
ending early due to an auxiliary field.
The dashed lines represent fixed points in the $(T/S)$ --
$n$ plane that result from models that do not end without an auxiliary
field.  In general, two-field models populate the same region as
one-field models and extend the $T/S \approx 0$ attractor to $n > 1$.
}
\label{fig:2fld}
\end{figure}

In our framework two-field models are models that would inflate
forever on their own.  We find
such models when the right hand sides of Eqs. (\ref{flowr},\ref{flown})
vanish prior to violating the slow-roll conditions.
When this happens, we obtain fixed points in the $T/S$ -- $(n-1)$ plane,
which are the most likely
values for $T/S$ and $(n-1)$ 50 $e$-folds prior to when the
second field ends inflation.
These points, shown in Fig.~\ref{fig:2fld}, populate
the region $T/S \approx 0$ and for $n>1$ along with the attractor line,
$T/S = -5(n-1)$.

It is also possible that a self-ending model has an auxiliary field that ends
inflation ``early''.  We treat this possibility by populating the
$(T/S)_{50}$ -- $(n-1)_{50}$ plane with the values of $T/S$ and $(n-1)$
at $N > 50$ for all one-field models.  We find that the two-field
models behave similarly to the one-field models.  The only
significant difference is that two-field models extend the $(T/S)_{50}
= 0$ attractor to $(n-1)_{50} > 0$ (see Fig.~\ref{fig:2fld}).

Finally, what about our taking $x^{\prime\prime}\simeq\,$
constant?   It can affect the relationship between
the initial and final values of $n-1$ and $T/S$ if $x^{\prime\prime}$
is large, since $x^{\prime\prime}$ need not be constant
(as is the case in some known models).
Since we have covered a wide range of initial values we would
expect that this fact would only slightly modify the
$(n-1)_{50}$ -- $(T/S)_{50}$ phase space; indeed, we have also
formulated the flow equations assuming $V^{\prime\prime\prime}/V =\,$
constant and obtain similar results.  Further, in the favored part
of the $(n-1)_{50}$ -- $(T/S)_{50}$ plane, $x^{\prime\prime}$ is small.

{\em Discussion.}  Prior to this work there was one guiding relation
for the inflationary observables:  $T/S = -5n_T$.  It has the
virtue of exactitude and can test the consistency of the scalar-field
inflationary framework, but it involves the power-law index of the gravity
wave perturbations, the most difficult inflationary
observable to measure.  By reformulating the equations governing
inflation, we have found generic relations
between $T/S$ and $(n-1)$:  Inflationary kinematics
constrain models to cluster along the lines $T/S =
-5 (n-1)$ and $T/S = 0$, with a forbidden region between
these two lines for $n<1$.  Large $n-1$ is possible, but at the
expense of a poor power-law for the density
perturbations (i.e., large $dn /d\ln k$), unless $T/S$ is
very small.  Further, our results support the view
that inflation generically predicts almost, but not exactly
scale-invariant density perturbations \cite{ST}.

These results provide practical guidance
to CMB experimenters and additional tests for inflation.
For example, if $n$ is found to be significantly greater than 1,
then a poor power-law is also expected unless $T/S \ll 1$.
If $n$ is found to be $>0.85$, then $T/S > 10^{-3}$ is likely,
which would makes prospects for detecting the gravity-wave
of inflation signature more favorable.

\paragraph{Acknowledgments.}  We thank Andrew Liddle and David Spergel
for valuable discussions and comments.  This work was supported by
the DoE (at Chicago and Fermilab) and by the NASA (at Fermilab
by grant NAG 5-7092).

\end{document}